\documentclass[prl,twocolumn,10pt,showpacs,aps]{revtex4-1}
\usepackage{graphicx,amsfonts,times,bm,amsmath, verbatim}

\begin{document}

\title{Magnetic Field Response and Chiral Symmetry of Time Reversal Invariant Topological Superconductors}

\author{Eugene Dumitrescu$^1$}
\author{Jay D. Sau$^2$}
\author{Sumanta Tewari$^{1}$}

\affiliation{
$^1$Department of Physics and Astronomy, Clemson University, Clemson, SC
29634\\
$^2$ Condensed Matter Theory Center, Department of Physics, University of Maryland, College Park, MD 20742
}
\begin{abstract}
We study the magnetic field response of the Majorana Kramers pairs of a
 one-dimensional time-reversal invariant (TRI) superconductors (class DIII)
with or without a coexisting chirality symmetry.
For unbroken TR \textit{and} chirality invariance the parameter regimes for nontrivial values of the ($\mathbb{Z}_2$) DIII-invariant and the ($\mathbb{Z}$) chiral invariant coincide. However, broken TR may or may not be accompanied by broken chirality, and if chiral symmetry is unbroken, the pair of Majorana fermions (MFs) at a given end survives the loss of TR symmetry in an entire plane perpendicular to the spin-orbit coupling field. Conversely, we show that broken chirality may or may not be accompanied by broken TR, and if TR is unbroken, the pair of MFs survives the loss of broken chirality. In addition to explaining the anomalous magnetic field response of all the DIII class TS systems proposed in the literature, we provide a realistic route to engineer a ``true" TR-invariant TS,
whose pair of MFs at each end is split by an applied Zeeman field in arbitrary direction. We also prove that, quite generally, the splitting of the MFs by TR-breaking fields in TRI superconductors is highly anisotropic in spin space, even in the absence of the topological chiral symmetry.
\end{abstract}

\maketitle

\textit{Introduction:}
Majorana fermions (MF), quantum zero energy particles with operators $\gamma=\gamma^{\dagger}$, are particles which can be identified
 with their own anti-particles \cite{Majorana,Wilczek,Frantz}. Because of their non-local topological nature they have been proposed
 to be useful in potential implementation of
 fault-tolerant quantum computation \cite{Kitaev-1D,Nayak_2008}.  Although MFs were first predicted in the context of high energy physics \cite{Majorana}, in low temperature systems they have been recently proposed as emergent quasi-particles in fractional quantum Hall systems \cite{Nayak-Wilczek,Read-Green}, 2D \cite{Read-Green} and 1D \cite{Kitaev-1D} spinless $p$-wave superconductors, naturally-occurring chiral $p$-wave superconductors such as strontium ruthenate \cite{Tewari-strontium},  and systems closely analogous to spinless $p$-wave superconductors such as heterostructures of topological insulators (TI) and $s$-wave superconductors \cite{Fu-Kane}, cold fermion systems with Rashba spin orbit coupling, Zeeman field, and an attractive $s$-wave interaction \cite{Zhang-Tewari,Sato-Fujimoto}, and also, most recently, heterostructures of  spin-orbit coupled semiconductor thin films \cite{Sau,Tewari-Annals,Long-PRB} and nanowires \cite{Long-PRB,Roman,Oreg} proximity coupled to $s$-wave superconductors and a Zeeman field. There have been recent claims of experimental observation of MFs in the semiconductor heterostructure, attracting considerable attention \cite{Mourik,Deng,Weizman,Rokhinson,Churchill,Finck,Appelbaum,Alicea,Leijnse,Beenakker,Stanescu}. In concurrent developments, recent work \cite{Schnyder_2008,Kitaev_2009,Ryu_2010} has established that the quadratic Hamiltonians describing gapped TI and topological superconductors (TS) can be classified into 10 distinct topological symmetry classes each of which can be characterized by a topological invariant.
 According to this classification, the experimentally investigated semiconductor heterostructures are in the topological symmetry class D with MFs protected by the particle-hole (p-h) symmetry $\Xi$ ($\Xi {\cal{H}}_{k} \Xi^{-1}=-{\cal{H}}_{-k}$ where ${\cal{H}}_{k}$ is the quadratic Bogoliubov-de Gennes (BdG) Hamiltonian).

Recently, a completely different class of 1D time reversal (TR) invariant topological superconductors (in class DIII) has been proposed where \textit{a pair} of MFs exist on each end of a quantum wire \cite{Law_12,Deng_12,Keselman_13,Zhang_13,Deng_13,Law_13,Flensberg_13,Nakosai_13}.
Even though a pair of zero energy states are localized at the same end, they are protected against hybridization (and acquiring finite energies) by the TR symmetry.
These systems are characterized by a DIII class $\mathbb{Z}_2$ invariant which takes a non-trivial value (indicating the existence of the pair of MFs at the same end) when the pairing potential has a negative sign on an odd number of Fermi surfaces each of which encloses a TR-invariant momenta \cite{Qi}. In Ref.~[\onlinecite{Zhang_13}] this is achieved by having a quantum wire with Rashba spin-orbit coupling proximity coupled to an unconventional $s_{\pm}$ wave superconductor with the superconducting pair potential $\Delta_{k}$ changing sign between the two Fermi surfaces.
 In the appropriate parameter regime the presence of the non-trivial value of the $\mathbb{Z}_2$  invariant implies that no perturbation respecting the TR symmetry can remove the MF pair from the wire ends without closing the bulk gap. Conversely, it is expected that perturbations that do break the TR symmetry can split the pair of MFs to higher energies by hybridization. Surprisingly, however, in all the examples of the DIII class TS systems proposed in the literature so far, the pair of MFs at each end continues to exist even in the presence of TR-breaking Zeeman fields in some directions in the spin space.

 In this paper we first show that 1D TR-invariant (TRI) TS systems in class DIII frequently possess a co-existing chiral symmetry allowing to define an integer $\mathbb{Z}$ topological invariant in addition to the DIII class $\mathbb{Z}_2$ invariant.
For unbroken TR invariance the parameter regimes for nontrivial values of the $\mathbb{Z}_2$ invariant and the $\mathbb{Z}$ invariant coincide, each indicating the presence of the Kramers pair of MFs. However, broken TR may not be accompanied by broken chirality, and if the chiral symmetry is unbroken, the pair of MFs at a given end can survive the loss of the TR symmetry. We show that, in this way, the existence of the chiral symmetry explains the persistence of the MFs even in the presence of magnetic fields in a \textit{plane} perpendicular to the spin-orbit coupling (the chiral symmetry was used earlier to explain the robustness of the zero modes to Zeeman fields in a specific direction in spin space in Rashba-coupled $d$-wave superconductors \cite{Sato_11,Law_12}.  Conversely, we find that broken chirality may also not be accompanied by broken TR symmetry, and in this case also the MFs remain un-split due to the persistence of the $\mathbb{Z}_2$ invariant. By numerically solving the appropriate BdG equations for a $s_{\pm}$-wave TRI superconductor we find that only those physical perturbations that simultaneously break both TR and chiral symmetries can hybridize the MFs and split them to finite energies.
In addition to solving the problem of the anomalous magnetic field response of all the DIII class TS systems proposed in the literature so far \cite{Law_12,Deng_12,Keselman_13,Zhang_13,Deng_13,Law_13,Flensberg_13} (we have checked that analysis similar to what follows applies to all of them), our results also provide a realistic route to engineer a ``true" TR-invariant topological superconductor
whose Majorana Kramers pairs are split by an applied Zeeman field in \textit{any} direction.
However, even in this case, and quite generally in TRI superconductors, we prove that the splitting of the MFs by TR-breaking Zeeman fields is highly anisotropic in spin space, and this can be taken as a ``smoking-gun" signature of TRI superconductivity and Majorana Kramers pairs.

\begin{figure}[ht]
\includegraphics[width=8cm]{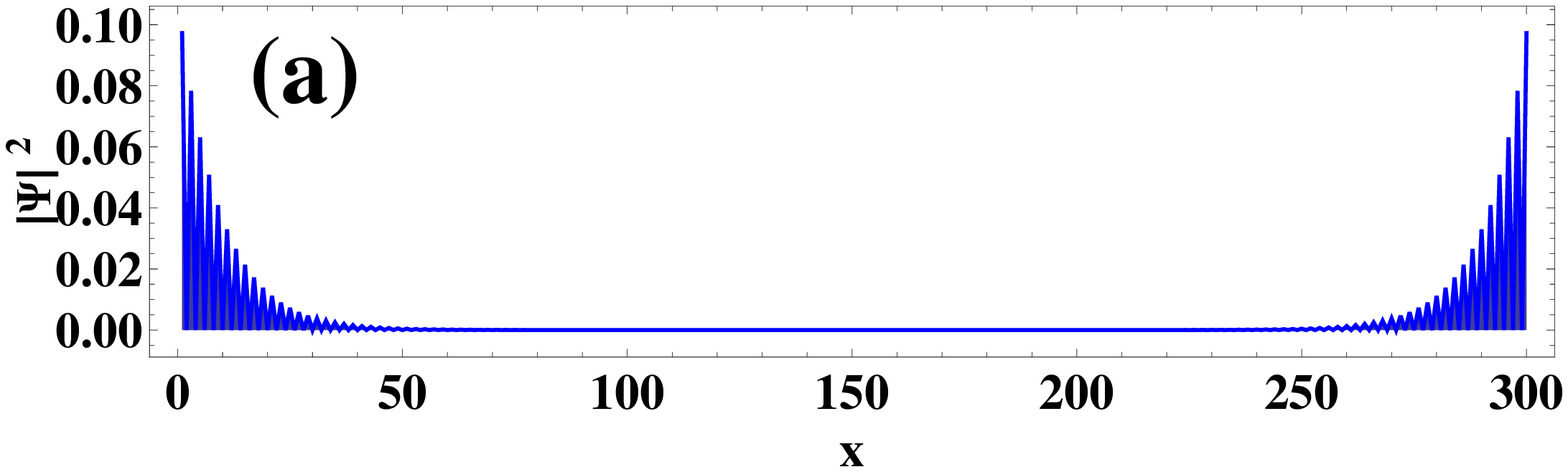}
\includegraphics[width=4cm]{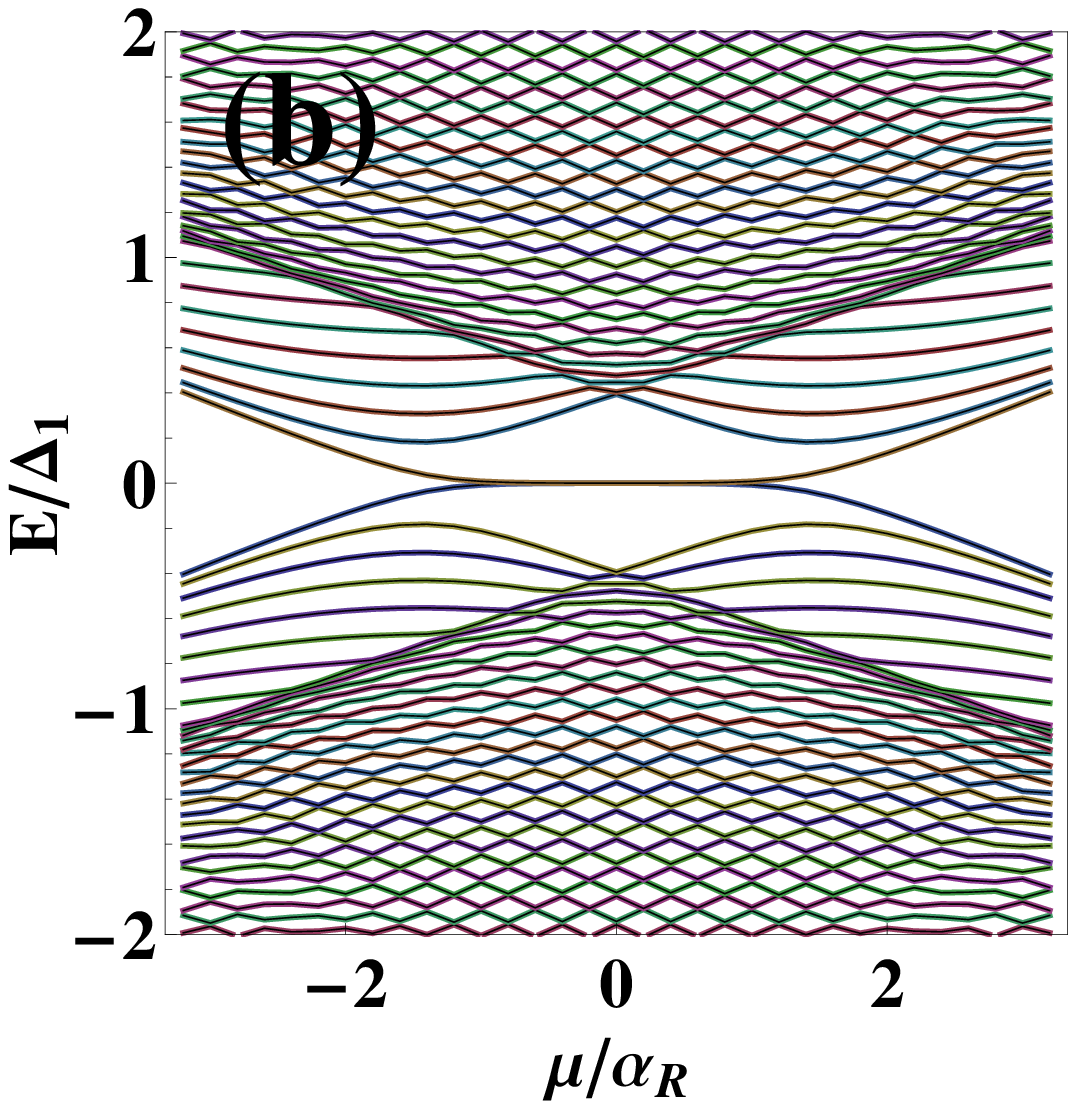} \includegraphics[width=4cm]{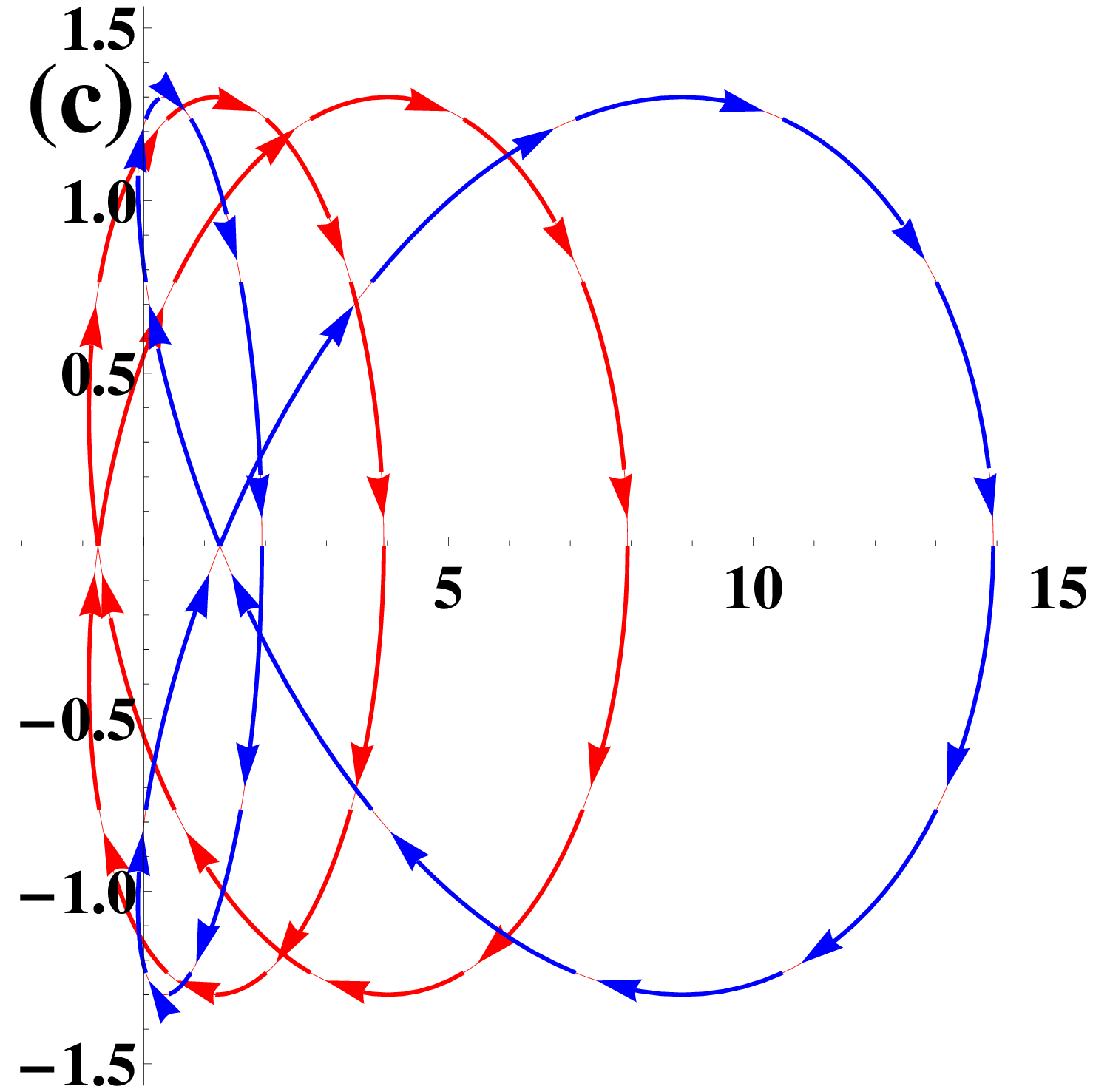}
\caption{\label{fig:winding} (Color online)
(a) Magnitude of the zero energy ground state BdG wavefunction along the length of the nanowire at $\mu=0$. The wavefunctions are localized about the endpoints and have equal particle and hole character satisfying the Majorana condition. (b) The low energy BdG spectrum as a function of chemical potential $\mu$. Each band is doubly degenerate, with 2 MFs at each end of the nanowire in the $|\mu|<2\alpha_R$ topologically non-trivial regime. (c) Parametric plot of Re$(Det(A(k)))$ and Im$(Det(A(k)))$ as the quasi-momentum $k$ is varied through the 1D Brillouin zone ($k\in[-\pi,\pi]$). The red curve ($|\mu|<2\alpha_R$, non-trivial regime) winds about the origin twice as $k$ is varied through the Brillouin zone yielding a chiral invariant $W=2$. Upon increasing the chemical potential, the blue curve ($|\mu|>2\alpha_R$ topologically trivial regime) winds about the origin 0 times indicating a decrease in the chiral invariant from $W=2$ to $W=0$ which is consistent with the disappearance of the MFs in (b) as well as the TR-invariant $\mathbb{Z}_2$ value.}
\end{figure}

\textit{Theoretical model and topological invariant:}
Let us consider a TR-invariant topological superconductor given by the following mean field BdG Hamiltonian $H=\sum_{k}\Psi_{k}^{\dagger} {\cal{H}}_{k} \Psi_{k}$ where,
\begin{equation}
\label{eq:BDG}
{\cal{H}}_{k}=(\epsilon_{k}-\mu)\sigma_{0}\tau_z+\alpha^{R}_{k}(\hat{a}\cdot\bm{\sigma})\tau_{z}+\Delta^{s}_{k}\sigma_{0}\tau_{x}
\end{equation}
where $\epsilon_{k}=-2t\cos(k)$ is the single particle kinetic energy, $\alpha^{R}_{k}=\alpha_{R}\sin k$ is the Rashba-type spin orbit interaction, and $\Delta^{s}_{k}=\Delta_{0}+\Delta_{1}\cos(k)$ represents the spin-singlet superconducting pair potential with a conventional $s$-wave order parameter ($\Delta_{0}$) and a $s_\pm$ \cite{Zhang_13} or $d$-wave component (\cite{Law_12}) ($\Delta_{1}$). The matrices $\sigma_{i}$,$\tau_{i}$ indicate spin $1/2$ Pauli matrices in the spin and the particle-hole spaces, respectively, and we have used the Nambu basis, $\Psi_{k}=(c_{k\uparrow},c_{k\downarrow},c_{-k\downarrow}^{\dagger},-c_{-k\uparrow}^{\dagger})^{T}$. The vector $\hat{a}$ indicates an arbitrary spin orbit coupling direction in the spin space. Selecting $\hat{a}$ along $\hat{y}$, Eq. (\ref{eq:BDG}) reduces to the 1D Hamiltonian used in \cite{Zhang_13}. We solve the corresponding lattice BdG equations for a finite wire setup, and find zero energy quasiparticle modes which satisfy the Majorana condition ($\gamma^{\dagger}=\gamma$) for the second quantized operators $\gamma_i$. The wavefunctions for the zero energy modes are localized at the wire ends which is illustrated in Fig. \ref{fig:winding} panel (a). The MFs exist for the chemical potential satisfying $|\mu|<2\alpha_R$, as seen in panel (b), corresponding to the range for which the $\mathbb{Z}_2$ TR invariant takes a non-trivial value \cite{Zhang_13}.

In addition to the particle hole symmetry, with the operator $\Xi=\sigma_y\tau_y{\cal{K}}$, the above Hamiltonian satisfies the TR condition $\Theta {\cal{H}}_{k} \Theta^{-1}={\cal{H}}_{-k}$ with the TR operator $\Theta=i\sigma_y\tau_0{\cal{K}}$ where ${\cal{K}}$ is the complex conjugation operator. For the parameter range $|\mu|<2\alpha_{R}$ the TR $\mathbb{Z}_2$ topological invariant has a non-trivial value and the system constitutes a topological superconductor with a pair of MFs localized at each end of the nanowire. Note however, that there exists an operator ${\cal O}=(\hat{a}\cdot\hat{y}+i(\hat{a}\times\hat{y})\cdot\bm{\sigma}){\cal K}$ with ${\cal O}^2=1$ which acts on the Hamiltonian similarly to the TR operator, ${\cal O} {\cal{H}}_{k} {\cal O}^{-1}={\cal{H}}_{-k}$. Due to the existence of ${\cal O}$ and the presence of $\Xi$, there exists a chiral operator which is simply the product $S_2={\cal O}\cdot \Xi=(\hat{a}\cdot\bm{\sigma}) \tau_y$. The operator $S_2$ anti-commutes with the Hamiltonian, $\{{\cal{H}}_{k},{\cal{S}}_2\}=0$ which can be easily shown by explicitly operating with $\Xi$ and then ${\cal O}$ on ${\cal{H}}_{k}$. Note that it is trivial to show that the operator ${\cal{S}}_1=\sigma_{0}\tau_{y}$ also anti-commutes with the BdG Hamiltonian as $\sigma_0$ ($\tau_y$) commutes (anti-commutes) with each term in Eq.~(\ref{eq:BDG}). However, invariants calculated with this operator (see below) are inconsistent with the behavior of the BdG spectrum and we note that $S_1, S_2$ form a two dimensional vector space of chiral operators anti-commuting with ${\cal{H}}_{k}$.

The existence of the $S_2$ symmetry allows us to compute an integer $\mathbb{Z}$ chiral invariant which counts the number of topologically protected MFs at each end of the wire. Because of the anti-commutation relation between the chirality operator and ${\cal{H}}_{k}$, the BdG Hamiltonian takes the off-diagonal form in the eigenbasis of $S_2$,
\begin{equation}
\label{eq:offdiag}
{\cal{H}}^{'}_{k}=U{\cal{H}}^{}_{k}U^{\dagger}=\left(\begin{array}{cc}
0 & A_k\\
A^{\dagger}_{-k} & 0
\end{array}\right).
\end{equation}
Where $U$ is the unitary transformation matrix between the two eigenbasis.
We write the determinant of $A_k$ in a complex polar form, $Det(A_k)=|Det(A_k)|\exp(i(\theta(k)))$.
For a fully gapped system, the topological invariant is given by the winding number $W$ \cite{Tewari_PRL_2012,Tewari_Minigap},
\begin{equation}
\label{eq:winding}
W=\frac{1}{2\pi i}\int_{-\pi}^{\pi}\frac{dz(k)}{z(k)}.
\end{equation}
The integer $W$ counts the number of times the angle $\theta(k)$ winds about the origin in the complex plane. This quantity is invariant under smooth deformations and cannot change without $|Det(A_k)|$ going to zero, indicating a gap closing and topological phase transition. As shown in Fig. \ref{fig:winding} panel (c) the winding number in the regime ($|\mu|<2\alpha_R$) is $2$, while in the topologically trivial regime the winding number becomes $0$. This corresponds exactly with the parameter regime for which the DIII class $\mathbb{Z}_2$ invariant takes a non-trivial value \cite{Law_12,Zhang_13}. Thus in the non-trivial topological regime the Hamiltonian given by Eq.~(\ref{eq:BDG}) is characterized by the coexistence of TR \textit{and} chiral symmetries. In the following discussion we investigate the behavior of the Majorana Kramers pair while breaking one symmetry and maintaining the other, and finnaly in the presence of terms which destroy both symmetries.

\begin{figure}[t]
\includegraphics[width=4cm]{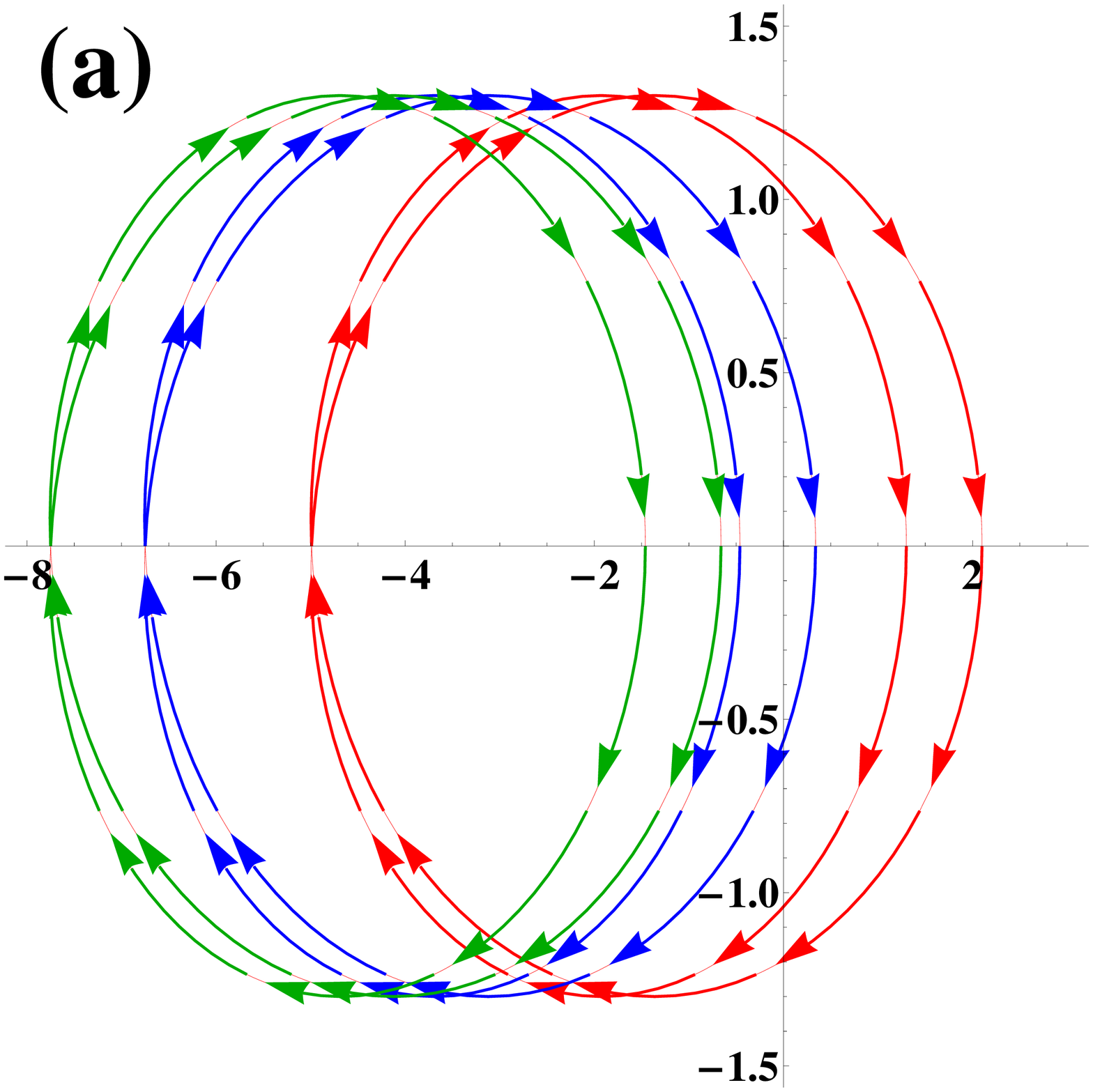} \includegraphics[width=4cm]{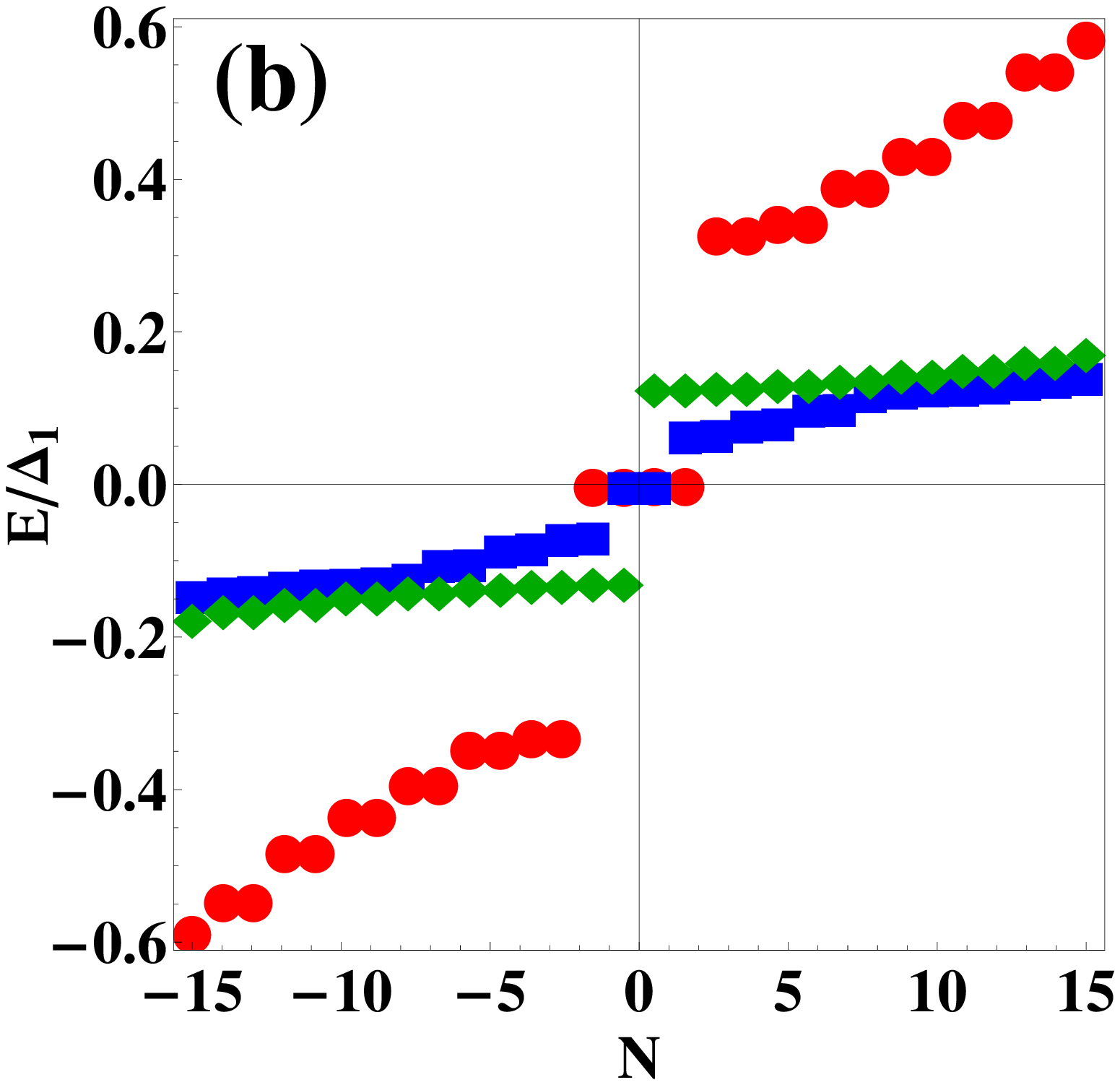}
\caption{\label{fig:TRbreaking} (Color online)
(a) Winding curves for the bulk BdG Hamiltonian given by Eqs. (\ref{eq:BDG},\ref{eq:Hz}) which preserves the chiral symmetry ($\bm{b} \perp \hat{a}$). The red, blue, and green curves yield chiral invariants $W=2,1,0$ connected by gap closures and topological phase transitions. (b) Low energy BdG quasiparticle spectrum corresponding to the winding curves in panel (a). The chiral invariant gives the number of topologically protected MF modes at each end of the nanowire.}
\end{figure}

\textit{TR-breaking perturbations maintaining chirality invariance:}
We consider the effect of TR-breaking perturbations such as Zeeman splitting in an arbitrary direction on the low energy BdG spectrum for a finite wire.  The Zeeman splitting is written as,
\begin{equation}
\label{eq:Hz}
{\cal{H}}^{Z}=V(\bm{b}\cdot\bm{\sigma})\tau_{0}.
\end{equation}
The total BdG Hamiltonian is a sum of Eq.~(\ref{eq:BDG}) and this term. Note that for $\bm{b} \perp \hat{a}$ this term anti-commutes with $S_2$, preserving the chiral symmetry of the system. As expected, the MFs are topologically protected and immune to splitting by Zeeman fields in the two directions (and on the entire plane formed by them) perpendicular to the spin-orbit field. Increasing the magnitude of the splitting removes one Fermi surface and the system is driven through a topological quantum phase transition into a state which is effectively a spin-less $p$-wave superconductor, characterized by one MF at each end of the wire. Further increasing the Zeeman splitting removes the second Fermi surface driving the system into a phase associated with no MFs. At each of these topological phase transitions the chiral invariant $W$ decreases by one from $W=2$ to $W=1$ and finally $W=0$ as seen in Fig. \ref{fig:TRbreaking} panel (a). Panel (b) shows the low energy BdG spectrum with parameters corresponding to those used for the winding curves. Thus, in the presence Zeeman splitting discussed above, it is appropriate to classify the system as a chiral topological superconductor with an integer $\mathbb{Z}$ invariant characterizing the topological phase. Note that a Zeeman splitting $\bm{b} \parallel \hat{a}$ commutes with $S_2$, breaking both TR and chiral symmetries. In this case the MFs do in fact couple as a result of the broken symmetries and the zero energy modes are split to finite energy (Fig.~[3], panel (a)).

\textit{Chirality breaking perturbations maintaining TR-invariance:}
Conversely, we consider terms in the BdG Hamiltonian which break the chiral symmetry while preserving TR-symmetry.
 For example, one may examine the response of the zero energy modes to the inclusion of a TR-invariant next nearest neighbor
spin-orbit coupling. In the absence of Zeeman splitting the total Hamiltonian is written as a sum of Eq. (\ref{eq:BDG}) and
\begin{equation}
\label{eq:Delta_p}
{\cal{H}}^{SO'}_{k}=\alpha'\sin(2 k)(\bm{c}\cdot\bm{\sigma})\tau_{z}.
\end{equation}
Here the vector $\bm{c}$ indicates the axis of the next-nearest neighbor spin-orbit coupling.
 Regardless of this direction Eq. \ref{eq:Delta_p} respects the TR condition, $\Theta {\cal{H}}_{k} \Theta^{-1}={\cal{H}}_{-k}$.
 If $\bm{c} \parallel \hat{a}$ the spin-orbit coupling has a fixed axis and the chiral $S_2$ symmetry is maintained as well. In this case, the results of the preceding section are applicable and the MFs are robust to the application of magnetic fields in the plane perpendicular to $\hat{a}$. However, for $\bm{c} \perp \hat{a}$, the extra spin-orbit coupling commutes with $S_2$, breaking the chiral symmetry. Physically,
the next-nearest neighbor spin-orbit coupling corresponds to the spin quantization axis being different at the different spin-split fermi points.
Such a difference in spin-polarizations between the two fermi points is natural in the limit where the fermi points are sufficiently
separated to allow a different sign of the pairing potential. Since such a difference in sign in the pair potential between the two spin-split Fermi points
is achieved in the topological regime ($|\mu| < 2\alpha_R$) \cite{Zhang_13} the next-nearest neighbor spin-orbit coupling is expected in this regime.
 Since the perturbation
${\cal{H}}^{SO'}$ does not break time-reversal symmetry, Kramers theorem dictates that Majorana fermions can only occur in
Kramers pairs and a single Kramers pair of Majorana modes at either of
the ends of the nanowire cannot be split by ${\cal{H}}^{SO'}$.

\textit{Zeeman splitting in a spin-nonconserving time reversal Hamiltonian:
 construction of a ``true" TR-protected topological superconductor:}
The addition of the ${\cal{H}}^{SO'}$ to break the chiral symmetry leads to a topological state in which spin is not conserved. (Note that, in the absence of
 such a term the spin operator $\hat{a}\cdot\sigma$ commutes with the Hamiltonian whereby this component of spin is conserved.)
We find that the addition of ${\cal{H}}^{SO'}$, which has
no chiral symmetry but is TR-invariant, leads to a Kramers pair of MFs that are split by Zeeman
splitting in all directions. However, as argued below,
even in this general time-reversal invariant case, where no chiral
symmetry is present, the splitting of the MFs at small Zeeman fields is still highly anisotropic in spin space, and this property
can be used as a diagnostic signature of Majorana-Kramers pairs in experiments.

 
 To understand the splitting in more detail, we consider the response of the MFs on the right end of
 the wire to an applied magnetic field.  To simplify the situation we assume that the MFs on the left end have been gapped out by
a localized magnetic field, which does not affect the right end of the wire. The splitting and crossing of the MFs on the right
end as a function of the magnetic field $V=\sum_{a=x,y,z}B_a \sigma_a$ can be analyzed by studying the Pfaffian of the BdG
Hamiltonian
\begin{equation}
Q(\bm B)=Pf(\sigma_y\tau_y H_{BdG}(\bm B)),\label{eq:Q0}
\end{equation}
where we have taken advantage of the fact that based on particle-hole symmetry of $H_{BdG}$, $\sigma_y\tau_y H_{BdG}$
is anti-symmetric and allows the definition of a Pfaffian. The Pfaffian $Q$ defined here is related to the Majorana number
defined by Kitaev \cite{Kitaev-1D}, and therefore corresponds to the fermion parity of the ground state. More importantly, Hermiticity of $H_{BdG}$
dictates that $Q(\bm B)$ is real and since $Q(\bm B)^2=Det(H_{BdG})$,
$Q(\bm B)$ can only change sign when an odd number of pairs of energy levels cross zero-energy.
 Therefore, in our system, where only the energy levels
on the right side can cross zero-energy when a small $\bm B$ is applied, the splitting and crossing of MFs is determined by $Q(\bm B)$.
The MFs are unsplit only when $Q(\bm B)=0$. Since the MFs are unsplit at $\bm B=0$, $Q(\bm B=0)=0$.
Applying the time-reversal operator $i\sigma_y K$ to Eq.~(\ref{eq:Q0}) we observe that $Q(\bm B)^2=Q(-\bm B)^2$.
This allows two choices $Q(\bm B)=\pm Q(-\bm B)$ where only the case with the  $-$ sign corresponds to having an odd number of
MFs that are split by the Zeeman field. This case corresponds to the
$\mathbb{Z}_2$ non-trivial TRI superconductor so that in this case $Q$ satisfies
\begin{equation}
Q(\bm B)=-Q(-\bm B).
\end{equation}
At small $\bm B$, one can expand $Q(\bm B)$ as
\begin{eqnarray}
Q(\bm B)&=&\sum_a [\rho_a B_a\{1+\sum_{b,c}\rho_{abc}B_b B_c(1+\sum_{d,f}(\rho_{abcdf}B_d B_f\nonumber\\
&+&\dots)\}],\label{eq:Q}
\end{eqnarray}
where $\rho_a$, $\rho_{abc}$, and $\rho_{abcdf}$ are coefficients that in principle can be determined in
perturbation theory.

\begin{figure}[ht]
\includegraphics[width=6cm]{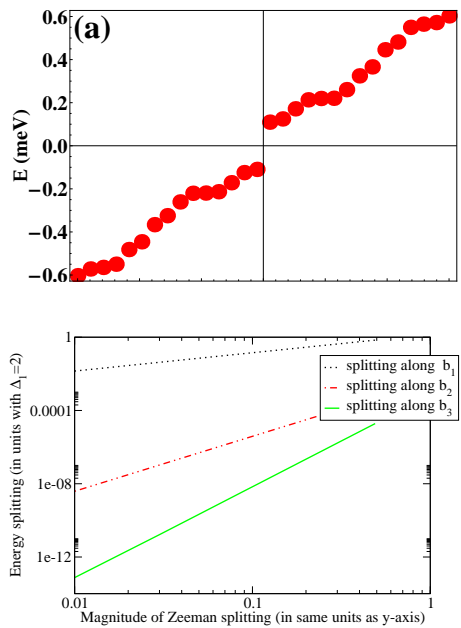}
\includegraphics[width=6cm]{energy_plot_TRinv.eps}
\caption{\label{fig:splitting} (Color online) (a) Energy splitting of the MFs in the presence of Zeeman splitting $V$ along $\hat{y}$ (along the direction of spin orbit direction) as described in Eq.~(\ref{eq:Hz}). (b) Energy splitting as a function of Zeeman field along three orthogonal directions $\bm b_{1,2,3}$
described in the text on a log-log scale are found (from the slope) to split as $V$,$V^3$ and $V^5$, where $V$ is the magnitude of the
Zeeman field. Thus, in the presence of next-nearest neighbor spin-orbit coupling described by Eq.~\ref{eq:Delta_p}, the MFs
split in every direction. Both axes are in units where the pairing potential
$\Delta_1=2$.}
\end{figure}

Using Eq.~\ref{eq:Q}, we note that there is a unique direction $\bm b_1$ parallel to $\bm\rho$, where the MFs split linearly
as long as $\bm B$ is not in the plane perpendicular to $\bm b_1$.
 This is similar to the chiral case where the Zeeman potential is applied along the axis of the spin-orbit
field. The existence of a unique axis with such a strong splitting of MFs is a general feature of the TR-invariant state.
In general the MF splitting when $\bm B$ is in the plane perpendicular to $\bm \rho$ is of order $B^3$. However, varying
$\bm B$ along the circle with $\bm B\cdot\bm\rho=0$ and $|\bm B|$ held fixed we conclude that
since $Q(\bm B)$ is odd in $\bm B$, the splitting must vanish at some point on the circle.
 Along this direction $\bm b_3$, the cubic coefficient also vanishes and the
splitting of the MFs is expected to be even slower i.e. of order $B^5$. The splitting in
the perpendicular direction $\bm b_2=\bm b_1\times \bm b_3$ is of order $B^3$.
By considering the nanowire with a next-nearest neighbor spin-orbit coupling with strength $\alpha'=\alpha$,
and calculating numerically the MF energies in the plane perpendicular to the direction where the MFs split linearly,
we see in Fig.~\ref{fig:splitting} that the MFs in this model split in all directions.

In conclusion, We study the magnetic field response of the Majorana kramers pairs of 
 one-dimensional TRI superconductors 
with or without a coexisting chirality symmetry.
 When the chiral symmetry is present, in addition to
TR, the Majorana Kramers pairs are topologically robust to TR-breaking fields on a plane perpendicular to the spin-orbit coupling.
Apart from explaining the anomalous magnetic field response of all the DIII class systems proposed in the literature \cite{Law_12,Deng_12,Keselman_13,Deng_13,Zhang_13,Law_13,Flensberg_13} (for instance, the chiral operator for the Hamiltonian in Ref. [\onlinecite{Keselman_13}] is $S_2= s_z \sigma_0 \tau_y$, where $\sigma_i (s_i)$ is a Pauli matrix in channel (spin) space), we provide a realistic route to engineer a ``true" TRI TS system
whose Majorana Kramers pairs are split by a Zeeman field applied in \textit{any} direction. However, even in this case, and quite generally in TRI superconductors, we prove that the splitting of the MFs is highly anisotropic in spin space, which can be used in experiments as a ``smoking-gun" signature of Majorana Kramers pairs.

This work is supported by NSF (PHY-1104527) and AFOSR (FA9550-13-1-0045) (E.D. and S.T). J.D.S. would like to acknowledge the University of Maryland, Condensed Matter theory center, and the Joint Quantum institute for start-up support.


\end{document}